\documentclass[12pt]{article}
\newcommand{\be}{\begin{equation}}
\newcommand{\ee}{\end{equation}}
\newcommand{\bea}{\begin{eqnarray}}
\newcommand{\eea}{\end{eqnarray}}
\begin{document}
\title{Driving Hamiltonian in a Quantum Searching Problem
  }
\author{Kazuto Oshima\thanks{E-mail: oshima@nat.gunma-ct.ac.jp}  \\ \\
\sl Gunma National College of Technology, Maebashi 371-8530, Japan }

\date{}
\maketitle
\begin{abstract}
We examine the driving Hamiltonian in the analog analogue of Grover's algorithm by Farhi and Gutmann.
For a quantum system with a given oracle Hamiltonian $E|w\rangle\langle w|$, we explicitly show that
while the driving Hamiltonian $E|s\rangle\langle s|$ optimally produces the state $|w\rangle$ from an initial state $|s\rangle$,
the driving Hamiltonian $E^{\prime}|s\rangle\langle s|(E^{\prime}\ne E)$ does not provide any speedup compared even with a classical computation.
\end{abstract}

PACS number(s):03.67.Lx\\
\newpage
In the ordinary paradigm for quantum computation\cite{Deutsch}, a computation is a sequence of elementary unitary transformations.
Recently, Farhi et.al.\cite{Farhi1} have advocated another type of quantum  computation, which has been applied to the NP-complete problem 3-SAT. In their method an initial ground state adiabatically develops in continuos time  obeying to a slowly varying Hamiltonian.  Choosing a suitable driving Hamiltonian we can obtain a preferable state with  high probability after a certain time.  Proceeding to this@adiabatic method, Farhi and Gutmann\cite{Farhi2} have proposed an analog analogue of the Grover's algorithm\cite{Grover}.   The adiabatic approximation is not necessary for this rather simple problem.  Since, it is important to design a driving Hamiltonian to shorten the time, in this brief report we examine the driving Hamiltonian
in \cite{Farhi2}. 

The problem we consider is to search an unknown state $|w \rangle$ in a unit sphere of $N$-dimensional complex vector space under a given Hamiltonian 
\begin{equation}
H_{w}=E|w \rangle \langle w|.
\end{equation}  
A restricted version of this problem is to find a state $|w \rangle$ from the set of $N$ orthonormal states $|a \rangle (a=1\sim N)$. This Hamiltonian is an oracle in the sense $H_{w}|a\rangle=\delta_{a,w}E|a\rangle$.
The procedure\cite{Farhi2} is first to choose an arbitrary initial state $|s \rangle$, or in the restricted version to start from the superposition state $|s \rangle={1 \over \sqrt{N}}\sum_{a=1}^{N}|a\rangle$.  Second we add the given Hamiltonian the following driving Hamiltonian   
\begin{equation}
H_{d}=E^{\prime}|s \rangle \langle s|.
\end{equation}
In  \cite{Farhi2} it has been shown by a rather indirect way that the case $E^{\prime}=E$ is optimal, which means that at $E^{\prime}=E$ the time the initial state $|s \rangle$ develops into $|\omega\rangle$ with the probability 1 almost attains the minimum value that is obtained by the general discussion.  
In this brief report we examine another possibility of the value of $E^{\prime}$ to shorten the time to obtain the state $|w\rangle$.  Even if the probability of obtaining the state $|w\rangle$ is less than 1 for a trial, we can obtain the state $|w\rangle$ with high probability by repeating trials.
It may be possible to realize total speedup by shortening the time for each trial.

The initial state develops under the total Hamiltonian $H=H_{w}+H_{d}$ as
$|\psi(t)\rangle=e^{-iHt}|s\rangle$. The Hamiltonian causes transition between
$|s\rangle$ and $|w\rangle$.   After some tedious calculations we obtain the following transition amplitude 
\begin{equation}
\langle w|\psi(t)\rangle=x\cos{\omega}t+i\cos(2\theta-\varphi)\sin{\omega}t,
\end{equation}
where $x=\langle w|s\rangle (0 \le x \le 1)$ which we can set real and nonnegative by choosing the phase of $|s\rangle$ properly, $\tan\varphi={\sqrt{1-x^{2}} \over x} (0 \le \varphi \le {\pi \over 2})$  and
\begin{equation}
\omega=E\sqrt{\left({1-\epsilon \over 2}\right)^{2}+{\epsilon}x^{2}},
\end{equation}
\begin{equation}
\tan\theta={{\epsilon}x\sqrt{1-x^{2}} \over {1-\epsilon \over 2}+{\epsilon}x^{2}-\sqrt{({1-\epsilon \over 2})^{2}+{\epsilon}x^{2}}}
\end{equation}
with $\epsilon={E^{\prime} \over E}$.   The angle $\theta$ increases from ${\pi \over 2}$ to ${\pi \over 2}+\varphi$ as $\epsilon$ increases from $0$ to $\infty$.
Especially, when $\epsilon=1$ we have $\theta={\pi \over 2}+{\varphi \over 2}$.
For $N$ large enough, $x$ asymptotically obeys the distribution $\sqrt{4N \over {\pi}}e^{-Nx^{2}}$. Hence, $x$ is expected to be $O({1 \over \sqrt{N}})$, and  in the following we set $x={1 \over \sqrt{N}}(\varphi \approx {\pi \over 2})$, which also is the value of $x$ in the restricted version. 

For $\epsilon=1$ we have $\cos^{2}(2\theta-\varphi)=1,\omega={E \over \sqrt{N}}$,
and at $t={2\sqrt{N} \over {\pi}E}$ the state $|s\rangle$ develops to $|w\rangle$,  which is the case in \cite{Farhi2}.  In contrast with this, for ${\epsilon}>1$ we have $\sin{\theta}={\epsilon \over \epsilon-1}{1 \over \sqrt{N}},\cos{\theta}=-1$ and for ${\epsilon}<1$ we have $\sin{\theta}=1,\cos{\theta}={\epsilon \over \epsilon-1}{1 \over \sqrt{N}}$.  Thus for $\epsilon \ne 1$ we have 
\begin{equation}
\cos^{2}(2\theta-\varphi)=\left({\epsilon+1 \over \epsilon-1}\right)^{2}{1 \over N}.
\end{equation}
This will mean that for $E^{\prime} \ne E$, even though the time $t={2 \over {\pi}\omega}$ that the amplitude $|\langle w|\psi(t)\rangle|$ reaches the maximum may become shorter, we need roughly $N$-times attempts to obtain the state $|w\rangle$.   Especially, in the limit $E^{\prime} \rightarrow \infty$, we need the same number of measurements  as the classical case.  Thus we cannot expect any speedup by the driving Hamiltonian with $E^{\prime} \ne E$. The optimal case $E^{\prime}=E$ seems very special.

\end{document}